\documentclass[pra,aps,twocolumn,showpacs,superscriptaddress,floatfix]{revtex4}

\usepackage{graphicx}
\usepackage{float}
\usepackage{pstricks}           
\usepackage{dcolumn}            
\usepackage{bm}                 
\usepackage{hyperref}           
\usepackage{hypernat}           
\usepackage[latin1]{inputenc}   
\usepackage{amsmath,amssymb}
\usepackage{version}
\DeclareGraphicsExtensions{.eps,.ps}

\newcommand{\bk}{{\bf k}}

\newcommand{\br}{{\bf r}}

\newcommand{\beqa}{\begin{eqnarray}}
\newcommand{\eeqa}{\end{eqnarray}}

\begin{document}

\title
{Electron-Ion Interaction Effects in Attosecond Time-Resolved
Photoelectron Spectra}
\author{C.-H. Zhang and U. Thumm}
\affiliation{Department of Physics, Kansas State University,
Manhattan, Kansas 66506, USA}
\date{\today}

\begin{abstract}

Photoionization by attosecond (as) extreme ultraviolet (xuv) pulses
into the laser-dressed continuum of the ionized atom is commonly
described in strong-field approximation (SFA), neglecting the
Coulomb interaction between the emitted photoelectron (PE) and
residual ion. By solving the time-dependent Sch\"{o}dinger equation
(TDSE), we identify a temporal shift $\delta \tau$ in streaked PE
spectra, which becomes significant at small PE energies. Within an
eikonal approximation, we trace this shift to the combined action of
Coulomb and laser forces on the released PE, suggesting the
experimental and theoretical scrutiny of their coupling in streaked
PE spectra. The initial state polarization effect by the laser pulse
on the xuv streaked spectrum is also examined.

\end{abstract}

\maketitle

\section{Introduction}
Using an xuv attosecond pulse to photoemit electrons from gaseous or
solid targets into the electric field of a synchronized delayed
femtosecond (fs) infrared (ir) laser pulse provides a powerful tool
for investigating ultrafast electron dynamics by recording
ir-laser-streaked xuv PE spectra~\cite{Krausz09}. For strong laser
fields and sufficiently fast PEs, streaked PE spectra are
conveniently described in SFA~\cite{Lewenstein94}, i.e., by ignoring
the interaction of the residual ion with the released PE. In this
case, subject only to the ir laser electric field, the propagation
of PEs can be described in terms of analytically known ``Volkov"
states~\cite{Zhang09}. This leads to a delay-dependent energy shift
$\delta E_{COE}^{SFA}(\tau)=-k A_L(-\tau)$ of the center-of-energy
(COE) $E_{COE}=\omega_X-|\varepsilon_B|$ in the PE spectrum, where
$\omega_X$ is the xuv-pulse central frequency, $\varepsilon_B$ the
binding energy in the initial bound state, $k$ the PE asymptotic
momentum, and $A_{L}$ the vector potential of the ir pulse. We use
atomic units except where stated otherwise and define the delay
$\tau$ between the centers of the xuv and ir pulse as positive if
the xuv pulse precedes the ir pulse.

The interpretation of sub-fs temporal shifts in streaked PE spectra
is a matter of current debate. For example, the recent
measurement~\cite{Cavalieri07} of a relative delay of $\approx 110
\pm 70$~as between the ir-streaked xuv photoemission from localized
4f core levels and delocalized conduction-band (CB) states of a
W(110) surface was understood in the original
ref.~\cite{Cavalieri07} and a subsequent theoretical
work~\cite{Kazansky09} as the difference $\delta t_{CB-4f}
=t_{CB}-t_{4f}$ between the arrival times of CB and 4f core PEs at
the surface. This interpretation is based on the assumption that the
ir pulse does not penetrate the surface, such that CB and 4f
electrons that are released at the same time $-\tau$ by the
absorption of an xuv photon get streaked only upon arrival at the
surface, producing the COE shifts $E_{COE}=-kA_L(-\tau+t_{CB})$ and
$-kA_L(-\tau+t_{4f})$, respectively, in the PE spectra. According to
this two-step explanation (photorelease followed by streaking), the
total emission probability $P_{tot} =\int dE P(E,\tau)$ from a given
initial state would not depend on $\tau$.

In contrast, our analysis of experimental streaked photoemission
data for a tungsten~\cite{Cavalieri07} and rhenium
surface~\cite{private} indicates that $P_{CB}(\tau)$ oscillates with
$A_L(\tau)$, with an amplitude of $\approx 10$\% of the average
value. Furthermore, the continuity of the wavefunction and its
derivative at the surface, implies that an intense fs ir pulse
affects the PEs inside the solid, even if the ir electric field were
prohibited from penetrating the surface~\cite{Varro98}. We have
shown that this observed temporal shift can be reproduced within the
SFA and interpreted it as an interference effect in the emission
from different lattice sites~\cite{Zhang09}, observing that the SFA
cannot account for relative temporal shifts in the emission from
different levels of isolated atoms. A classical transport simulation
including the effect of (in)elastic collisions of released PEs with
tungsten cores on the propagation of PE inside the solid leads to
$\delta t_{CB-4f}=33$~as~\cite{Lemell09}. Thus, different
models~\cite{Cavalieri07,Kazansky09,Zhang09,Lemell09,Baggesen09}
strongly deviate with regard to the assumed attenuation of the
streaking ir electric field $E_{L}(t)$ inside the solid, ranging
from no penetration into the surface~\cite{Kazansky09} to
penetration depths of 30~\cite{Zhang09} and 85 layer
spacings~\cite{Lemell09} or larger than the electron mean free
path~\cite{Baggesen09}. The detailed modeling of the (relative)
delay in the photoemission from metal surfaces is further
complicated by the complex band structure and the ensuing difficulty
in assigning a group velocity to the motion of PE wave packets
inside the dispersive conduction band~\cite{Cavalieri07}, surface
charge accumulation, and the general concern that static
band-structure calculations and the assumption of an instantaneous
plasmon response (i.e., static image charge interactions) are
invalid at the as time scale. These shifts may be of particular
importance in the interpretation of streaking spectra for complex
targets, such as metals, and emphasize the need for more detailed
studies of streaked photoemission spectra.

It is of fundamental importance to first understand all
contributions to this temporal shift for simple systems. In this
work, we focus on the effect of simultaneous ir laser pulse and
Coulomb interactions on streaked photoemission spectra  from the
prototypical ground state of a one-dimensional hydrogen atom. This
Coulomb-laser coupling effect was first investigated by Kroll and
Watson~\cite{Kroll73} in their study of laser-assisted atomic
scattering. It also affects the spectra of high harmonic generation,
multiphoton ionization, and laser-assisted xuv photoionization. For
example, the Coulomb interaction causes the xuv streaked PE spectra
to be right-left asymmetric~\cite{Smirnova08}. In RABITT
measurements (Reconstruction of Attosecond Beating by Interference
of Two photon Transition), these simultaneous ir laser pulse and
Coulomb interactions induce a so-called atomic phase which shifts
sideband intensities as a function of the delay between the xuv
pulse train and the ir
pulse~\cite{Veniard96,Toma02,Mauritsson05,Varju05}.

In this work, we demonstrate, numerically and analytically, how the
{\em coupling} of the ir laser pulse and the final-state Coulomb
interaction of the PE with the residual ion gives rise to a
significant temporal shift $\delta \tau$ in the COE of streaked PE
spectra with respect to those approximated in SFA. As we will show,
inclusion of this Coulomb-laser (CL) coupling alters both, the
amplitude and phase of the COEs in streaked PE spectra, leading to a
COE shift $\delta E_{COE}^{CL}(\tau)=-K A_L(\tau-\delta\tau)$ with
an oscillation amplitude $K>k$. Thus, $\delta\tau $ and the
streaking amplitude ratio $K/k$ i) help to reveal details of the PE
dynamics including the combined interaction of Coulomb and laser
forces and ii) converge to their SFA limits, $0$ and $1$,
respectively, at sufficiently large PE energies.

\begin{figure}[t]
\begin{center}
\includegraphics[width=1.0\columnwidth,keepaspectratio=true,
draft=false]{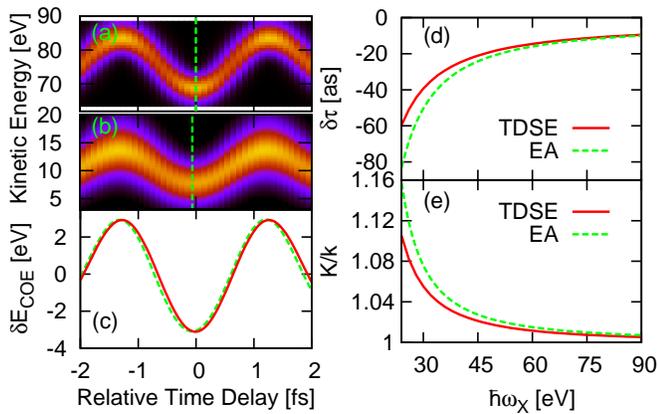}  \vspace{-6mm} \caption{(Color online)
Streaked photoemission from 1D model hydrogen atoms. TDSE
calculations for xuv pulses with (a) $\hbar \omega_X=90$ and (b)
$25$~eV. (c) Corresponding centers-of-energy $\delta E_{COE}(\tau)$
for $\hbar \omega_X=90$ (solid line) and $25$~eV (dashed line).
To facilitate the identification of the relative temporal shifts
$\delta\tau$, $\delta E_{COE}(\tau, \hbar \omega_X=90$~eV) is
normalized to the $\hbar \omega_X=25$~eV result. (d) $\delta\tau$
and (e) oscillation amplitude relative to the SFA for TDSE (full
line) and eikonal approximation (dashed line) calculations.
\label{fig:coe_exuv} } \vspace{-6mm}
\end{center}
\end{figure}

We numerically solve the TDSE, including the electron-proton
interaction, and then compare our results for $P(E,\tau)$ with SFA
and eikonal approximation (EA) calculations for a large range of PE
kinetic energies. Strongly dependent on $\omega_X$, we find temporal
shifts $\delta\tau$ of more than 50~as. Fig.s~\ref{fig:coe_exuv} (a)
and (b) show the TDSE ir-streaked PE spectra $P(E,\tau)$ for
gaussian xuv pulses of length $\tau_{X}=300$~as with
$\hbar\omega_X=90$ and $25$~eV, respectively. The ir pulse is also
assumed to be a gaussian and has a peak intensity of
$I_{L}=2\times10^{12}$~W/cm$^2$, a carrier frequency
$\omega_{L}=1.6$~eV/$\hbar$, and a pulse length $\tau_L=5$~fs. These
spectra are shifted by $\delta\tau=60$~as, which becomes apparent in
the corresponding COE shifts $\delta E_{COE}(\tau) $ in
Fig.~\ref{fig:coe_exuv} (c). The solid curves in
Fig.s~\ref{fig:coe_exuv} (d) and (e) show $\delta \tau$ and the
ratio $K/k$ of streaking-oscillation amplitudes for a large range of
$\omega_X$. Within an EA
approach~\cite{Joachain83,Gersten75,Smirnova08}, we can trace
(details will be given further below) this $\omega_X$-dependent
temporal shift and the oscillation amplitude enhancement to the CL
coupling in the PE final state (dashed curves in
Fig.s~\ref{fig:coe_exuv} (d) and (e)). $\delta\tau$ and $K/k$, for
the TDSE and EA calculations, converge at large $\omega_X$ to their
respective SFA limits $\delta\tau=0$ and $K/k=1$ due to the
diminishing influence of the residual ion's Coulomb force at
increasing PE energies. Thus, $\delta\tau$ and $K/k-1$, are measures
for the combined action of the Coulomb and laser force on the PE
relative to the action of the ir laser force alone. Since $\delta
\tau<0$, the attractive Coulomb force does not delay the PE
emission, as one might intuitively expect. We also note that $K>k$
reveals a Coulomb-enhancement effects that is reminiscent of the
Coulomb potential's infinite range leading to well-understood
``Coulomb-Cusps" in energy-differential collision-induced PE
spectra~\cite{Thumm92}.

This article is organized as follows. In Sec. \ref{sec:tdse}, we
present numerical results based on the TDSE. In Sec.
\ref{sec:eikonal}, we adopt an EA to take into account the
simultaneous ir and Coulomb interactions of the PEs, and compare our
EA and TDSE results. In Sec. \ref{sec:polarization}, we examine the
effect of the polarization of the initial state on the streaked
spectrum. We conclude in Sec. V. In the appendix, we show that,
within the eikonal approximation, the obtained atomic phase in
RABITT is identical to the relative temporal shift in the streaked
PE spectrum.

\section{Time-Dependent Schr\"{o}dinger Equation for Streaking}
\label{sec:tdse}

The exact wavefunction of the one-dimensional model atom interacting
with the ir and xuv pulse is determined by the TDSE (in the length
gauge)
\begin{align}
\label{eq:tdse} i\frac{\partial}{\partial
t}\Psi(x,t)&=\left[-\frac{1}{2}\frac{d^2}{dx^2}+U(x)+V(x,t)\right]
\Psi(x,t),
\end{align}
where $U(x)$ is the Coulomb potential,
$V(x,t)=x\left[E_{L}(t)+E_X(t)\right]$ the interaction with the ir
and xuv pulse, and $E_{L(X)}$ the electric field of the ir (xuv)
pulse. Assuming single-photon ionization in a sufficiently weak xuv
pulse, and after splitting the exact wave function for the atom in
the combined xuv and ir electric fields according to
$\Psi(x,t)=\psi_g(x,t)+\delta\psi(x,t)$, (\ref{eq:tdse}) can be
replaced by two coupled equations~\cite{Kazansky07}
\begin{align}
\label{eq:init} i\frac{\partial}{\partial
t}\psi_g(x,t)&=\left[-\frac{1}{2}\frac{d^2}{dx^2}+U(x)+xE_{L}(t)\right]
\psi_g(x,t),
\displaybreak[0]\\
\label{eq:ex} i\frac{\partial}{\partial
t}\delta\psi(x,t)&=\left[-\frac12\frac{d^2}{dx^2}+U(x)+xE_{L}(t)\right]
\delta\psi(x,t)
\displaybreak[0]\nonumber\\
&\ \ \ +xE_X(t+\tau)\psi_g(x,t).
\end{align}
Equation (\ref{eq:init}) determines the evolution (polarization) of
the initial state in the ir field and (\ref{eq:ex}) the generation
of PE wave packets by the xuv pulse and their evolution in the ir
field. The electric fields $E_{L(X)}=-\partial A_{L(X)}(t)/\partial
t$ of the ir (xuv) pulses are derived from the vector potentials
$A_{L(X)}(t)=A_{L(X),0}\cos(\omega_{L(X)}t)e^{-2\log2
t^2/\tau^2_{L(X)}}$. Since $E_{L(X)}(t\rightarrow\pm\infty)=0$,
equations (\ref{eq:init}) and (\ref{eq:ex}) are subject to the
initial conditions
$\psi_g(x,t\rightarrow-\infty)=\psi(x)e^{-i\varepsilon_Bt}$ and
$\delta\psi(x,t\rightarrow-\infty)=0$. The ground-state initial wave
function  $\psi(x)$ and energy $\varepsilon_B$ are obtained from
\begin{align}
\label{eq:gs}
\varepsilon_B\psi(x)&=\left[-\frac{1}{2}\frac{d^2}{dx^2}+U(x)\right]\psi(x).
\end{align}
We solve (\ref{eq:init})-(\ref{eq:gs}) numerically by wave-packet
propagation for times $|t|\le 2.5\tau_L$ with a step size $\Delta
t=0.2$ on a spatial grid with $|x| \le 2000$ and spacing $\Delta x
=0.25$. Assuming free-electron dispersion, $E=\frac12k^2$, we
calculate the ir-assisted xuv photoemission probability
\begin{align}
\label{eq:probtau}
P(E,\tau)=\left|\delta\tilde{\psi}(k,\tau,t\rightarrow\infty)\right|^2
\end{align}
and the corresponding COE
\begin{align}
E_{COE}(\tau)=\frac12\int dk\left|k \,
\delta\tilde{\psi}(k,\tau,t\rightarrow\infty)\right|^2/P_{tot}(\tau),
\end{align}
where $\delta\tilde{\psi}(k,\tau,t)$ is the Fourier transform of
$\delta\psi(x,t)$, and the total emission probability is
\begin{align}
P_{tot}(\tau)=\int dk
\left|\delta\tilde{\psi}(k,\tau,t\rightarrow\infty)\right|^2.
\end{align}

We model the target atom based on the soft-core Coulomb potential
\begin{align}
U(x)=-\frac{1}{\sqrt{x^2+a^2}},
\end{align}
and adjust the parameter $a=\sqrt{2}$ to the ground state binding
energy $\varepsilon_B=-13.6$~eV of the hydrogen atom.  We refer to
the exact solution of (\ref{eq:init})-(\ref{eq:gs}) as ``TDSE
result" and retrieve the SFA results by ignoring $U(x)$ in
(\ref{eq:ex}). The comparison of TDSE and SFA results is shown in
Fig.s~\ref{fig:coe_exuv}(a)-(c). By dropping the laser interaction
$xE_L(t)$ in (\ref{eq:init}), we verified numerically that for the
given parameters the polarization of this initial state by the ir
pulse can be neglected~\cite{Kazansky07,Baggesen09}. The initial
state polarization effect on the streaked xuv PE spectrum is further
discussed in Sec.~\ref{sec:polarization}.


\section{Eikonal Approximation}
\label{sec:eikonal}

In order to trace the influence of the combined action of Coulomb
potential and ir pulse on the PE, we write the PE wave function as
\begin{align}
\label{eq:wave} \psi_k(x,t)=a_k(x,t)
e^{i[k+A_L(t)]x-ik^2t/2+iS_k(x,t)}
\end{align}
with a local phase $S_k(x,t)$. The real amplitude $a_k$ is not
important for the present investigation. In SFA, the phase
$S_k(x,t)$ is given by the Volkov phase
\begin{align}S_k^{SFA}(t)=k\int_t^{\infty} dt'
A_L(t')
\end{align}
and independent of $x$. In EA, and {\em without} the ir field, the
phase accumulated by the PE during its propagation in $U(x)$ from
the location $x$ at time $t$ to the electron detector is calculated
along the free-electron classical trajectory
$x^\prime(t',t,x)=x+k(t'-t)$~\cite{Joachain83}
\begin{align}
\label{eq:EA} S^{C}_k(x)=\int_{t}^{\infty}dt'
U[x'(t',t,x)]=\frac{1}{k}\int_x^{\infty}dx'U(x').
\end{align}
In the presence of the ir field, the free-electron classical
trajectory is modified by a laser-induced drift
\begin{align}
\Delta x(t,t')&=\int_t^{t'}dt'' A_L(t'')
\end{align}
to become
\begin{align}
x_L(t',t,x)=x^\prime(t',t,x)+\Delta x(t,t').
\end{align}
Replacing $x'$ with $x_L$ in (\ref{eq:EA}), we obtain the CL
phase~\cite{Gersten75,Smirnova08}
\begin{align}
S^{CL}_k(x,t)=\int_{t}^{\infty}dt' \, U[x_{L}(t',t,x)],
\end{align}
and an eikonal approximation to the local phase in (\ref{eq:wave})
\begin{align}
\label{eq:EAA} S^{EA}_k(x,t)=S_k^{SFA}(t)+ S^{CL}_k(x,t).
\end{align}

In typical streaking experiments and for this study, the ir
intensity ($\sim 10^{12}$ W/cm$^2$) is low enough for $\Delta
x(t,t')$ being a small deviation from $x^\prime(t',t,x)$. We thus
expand $S^{CL}_k$ about $x^\prime(t',t,x)$  and obtain to first
order in $\Delta x(t,t')$
\begin{align}
\label{eq:Gk} S^{CL}_k(x,t)&=S^{C}_k(x)-\int_{t}^{\infty}dt'
F[x^\prime(t',t,x)]\Delta x(t,t'),
\end{align}
with the Coulomb force
\begin{align}
F[x^\prime(t',t,x)]=-\frac{\partial
U[x^\prime(t',t,x)]}{\partial  x^\prime}.
\end{align}
The first term in (\ref{eq:Gk}) is the laser-free eikonal Coulomb
phase. This term is independent of time as we explicitly indicate in
(\ref{eq:EA}). As we should show below this phase does not induce
any temporal shift in the streaked xuv spectrum, but changes the
transition probability. The second term in (\ref{eq:Gk}) includes
Coulomb scattering of the PE while it absorbs or releases ir
photons~\cite{Gersten75}. It is proportional to the ir vector
potential and causes a temporal shift in the streaked spectrum.

\begin{figure}[t]
\begin{center}
\includegraphics[width=1.0\columnwidth,keepaspectratio=true,
draft=false]{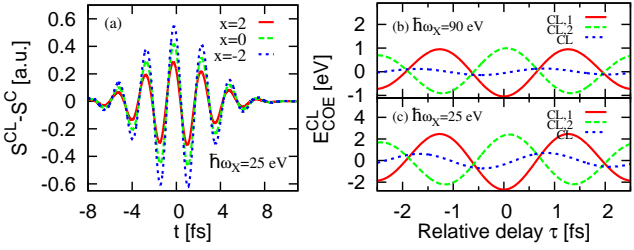}  \vspace{-6mm} \caption{(Color online) (a)
Time evolution of the CL coupling phase (see text).
Contribution to the PE streaking $\delta E^{EA}_{COE}(\tau)$ at
$x=0$ due to final state CL coupling  at (b) $\hbar\omega_X=90$ and
(c) 25~eV. \label{fig:eikonal} }\vspace{-6mm}
\end{center}
\end{figure}

The numerical results for (\ref{eq:Gk}) in Fig.~\ref{fig:eikonal}
(a) resolve the spatial contributions to the CL coupling phase
$S^{CL}_k(x,t)-S^{C}_k(x)$. Keeping in mind that the EA is designed
for short PE de-Broglie wavelengths~\cite{Joachain83}, we confirmed
by comparison with full TDSE results (not shown) that
$S^{CL}_k(x,t)$ remains appropriate down to $k=1$, i.e., a PE energy
of $\approx 14$~eV~\cite{Smirnova08}. This supports the validity of
the EA for the range of PE kinetic energies in
Fig.s~\ref{fig:coe_exuv} and \ref{fig:eikonal}.

The transition amplitude for xuv photoemission from the initial
state $\psi_i$ to the final state $\psi_k$,
\begin{align}
\label{eq:tmatrix} T_k(\tau)=-i\int\!dt \,
\langle\psi^*_k(t)|xE_{X}(t+\tau)|\psi_i(t)\rangle,
\end{align}
provides the PE probability $P(E=k^2/2,\tau)=|T_k(\tau)|^2$ as an
alternative to (\ref{eq:probtau}). Neglecting the laser distortion
of the initial state (using $\psi_i(x,t) \approx
\psi_i(x)e^{-i\varepsilon_Bt}$) and employing the EA-approximated PE
wave function for $\psi_k(x,t)$, we obtain
\begin{align}
T^{EA}_k(\tau)=&-i\int\!dt\,dx \, a(x,t)e^{-i[k+A_L(t)]x}\, x\psi(x)
\nonumber\\
&\times E_X(t+\tau)e^{-iS^{EA}_k(x,t)}e^{-i(\varepsilon_B-k^2/2)t}.
\label{eq:T}
\end{align}
The COE of the spectrum for a free PE would be
$E_{COE}=k^2/2=\omega_X-|\varepsilon_B|$. The (local) energy shift
caused by the ir field and CL coupling in EA is given by
\begin{align}
\delta E^{EA}_{COE}(x,t)=\partial S^{EA}_k(x,t)/\partial t,
\end{align}
and does not depend on the time-independent laser-free eikonal phase
$S^C_k(x)$ in (\ref{eq:Gk}). For sub-fs xuv pulses, contributions to
the time integral (\ref{eq:T}) mainly arise near the center of the
xuv pulse at $t=-\tau$. Approximating $\delta
E^{EA}_{COE}(x,\tau)\approx\partial S(x,t=-\tau)/\partial t$, we
obtain
\begin{align}
\label{eq:EA_Ecoe} \delta E^{EA}_{COE}(x,\tau)=- k A_L(\tau) +
E_{COE}^{CL,1}(x,\tau) +E_{COE}^{CL,2}(x,\tau),
\end{align}
where
\begin{align} \label{eq:Ec1}
E_{COE}^{CL,1}(x,\tau)&=\frac{U(x)}{k}A_{L}(\tau),\\
\label{eq:Ec2}
E_{COE}^{CL,2}(x,\tau)&=-\frac{1}{k}\int_{x}^{\infty}dx'F(x')A_{L}
\left(\frac{x'-x}{k}-\tau\right)
\end{align}
are the two contributions to the CL shift,
$E_{COE}^{CL}=E_{COE}^{CL,1}+E_{COE}^{CL,1}$, shown in
Fig.s~\ref{fig:eikonal} (b) and (c) at $x=0$ for $\hbar \omega_X=90$
and $25$~eV.

\begin{figure}[t]
\begin{center}
\includegraphics[width=1.0\columnwidth,keepaspectratio=true,
draft=false]{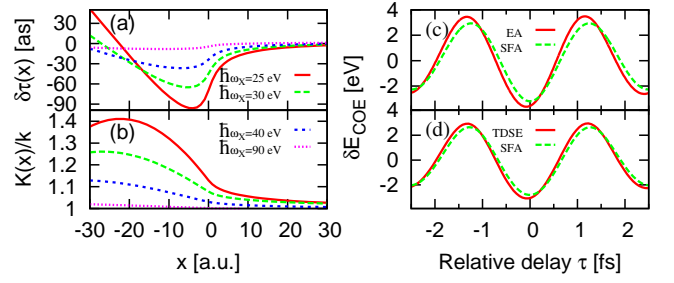}  \vspace{-6mm} \caption{(Color online) (a)
Local temporal shift $\delta\tau(x)$, and (b) local oscillation
amplitude $K(x)$ induced by the CL interaction in EA. Comparison of
the streaked COEs in SFA for $\hbar \omega_X=25$~eV with streaking
energies (c) in EA at $x=0$ and (d) from full TDSE calculations. The
PE is assumed to move to the right ($k>0$).
\label{fig:eikonal_exuv25}} \vspace{-6mm}
\end{center}
\end{figure}

According to (\ref{eq:Ec1}) and (\ref{eq:Ec2}),
$E_{COE}^{CL}(x,\tau)$ is proportional to $1/k$, while $\delta
E_{COE}^{SFA}(\tau)$ is proportional to $k$. Therefore, the CL
coupling effect decreases for increasing PE kinitic energies. As
shown in Fig.s~\ref{fig:eikonal} (b) and (c), the cancelation
between $E_{COE}^{CL,1}$ and $E_{COE}^{CL,2}$ becomes stronger and
further reduces the CL coupling with increasing $k$. Note that
$E_{COE}^{CL,1}(x,\tau)$ mainly increases the oscillation amplitude
of the COE in SFA, while $E_{COE}^{CL,2}(x,\tau)$ changes the
oscillation amplitude {\em and} induces a phase shift. We can thus
introduce a local temporal shift $\delta\tau(x)$ (relative to the
SFA phase) and  a local oscillation amplitude $K(x)$ by rewriting
(\ref{eq:EA_Ecoe}) as
\begin{align}
\delta E_{COE}^{EA}(x,\tau)= K(x)A_{L}[\tau-\delta\tau(x)].
\end{align}
Fig.s \ref{fig:eikonal_exuv25} (a) and (b) show $\delta\tau(x)$ and
$K(x)$ as functions of $x$ for different $\omega_X$. $\delta\tau(x)$
can be positive or negative. The actual shift $\delta\tau$ in the
streaking spectrum is obtained by spatial integration according to
(\ref{eq:T}). However, as shown in Fig.~\ref{fig:eikonal_exuv25} (c)
and (d), $\delta E_{COE}^{EA}(x=0,\tau)$ agrees well with the TDSE
result, since the initial wave function $\psi(x)$ is localized at
$x=0$. Similarly, we find that the full TDSE results for
$\delta\tau$ (solid line in Fig.~\ref{fig:coe_exuv} (d)) and $K/k$
(solid line in Fig.~\ref{fig:coe_exuv} (e)) agree well with the EA
results $\delta\tau(x)$ (dashed line in Fig.~\ref{fig:coe_exuv} (d))
and $K(x)/k$ (dashed line in Fig.~\ref{fig:coe_exuv} (e)) evaluated
at $x=0$. This justifies approximating $\Delta
S_k^{EA}(x,t)=S^{EA}_k(x,t)-S^{C}_k(x)$ in (\ref{eq:EAA})
\begin{align}
\label{eq:Gk2} \Delta S^{EA}_k(x,t)&\approx
S_k^{SFA}(t)-\int_{t}^{\infty}dt'
F[x^\prime(t',t,x=0)]\Delta x(t,t') \nonumber\\
&\approx-\frac{KE_L(t+\delta\tau)}{\omega_L^2},
\end{align}
where, in the second line, the slow varying envelop approximation is
used.

\begin{figure}[t]
\begin{center}
\includegraphics[width=1.0\columnwidth,keepaspectratio=true,
draft=false]{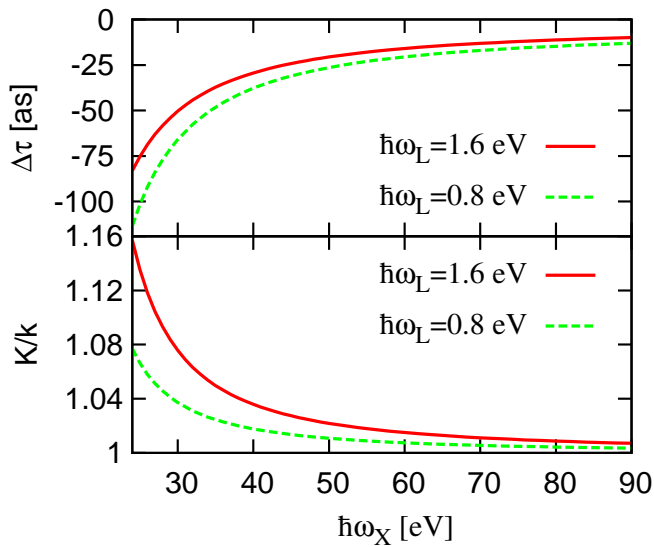} \vspace{-6mm} \caption{(Color online)
Comparison of eikonal temporal shift $\delta\tau$ and oscillation
amplitude ratio $K/k$ at two $\omega_L$. \label{fig:coe_omega0.8}}
\vspace{-6mm}
\end{center}
\end{figure}

We notice from (\ref{eq:EA_Ecoe})-(\ref{eq:Ec2}) that the three
contributions to $\delta E^{EA}_{COE}(x,\tau)$ are equally
proportional to the ir electric field amplitude. Therefore,
$\delta\tau$ and $K/k$ do not depend on the intensity of the ir
pulse. However, reducing $\omega_L$, $|\delta\tau|$ increases and
$K/k$ decreases (Fig.~\ref{fig:coe_omega0.8}). This is consistent
with (\ref{eq:Ec2}): at smaller $\omega_L$, $A_L$ oscillates slower,
leading to less cancelation in the time integral and thus to larger
$\delta\tau$. Simultaneously, stronger cancelation between $\delta
E_{COE}^{CL,1}$ and $\delta E_{COE}^{CL,2}$, results in smaller
$K(x)$.

\section{Polarization of the Initial State by the IR Pulse}
\label{sec:polarization}

The effect of initial state polarization by the ir pulse on the
streaked xuv photoemission spectrum has been addressed
previously~\cite{Kazansky07,Smirnova06,Baggesen10}. In this section,
we analyze how it affects the temporal shift $\delta\tau$ and the
oscillation amplitude ratio $K/k$. We find that the significance of
the initial-state polarization depends on whether or not the laser
unperturbed initial state is energetically isolated from other
levels.

\begin{figure}[t]
\begin{center}
\includegraphics[width=1.0\columnwidth,keepaspectratio=true,
draft=false]{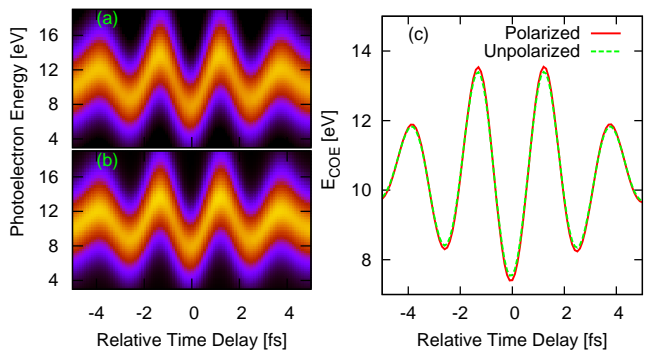}  \vspace{-6mm} \caption{(Color online)
Polarization effect of the initial state on ir-streaked
photoemission from the ground state of 1d model hydrogen atoms for
$\hbar\omega_X=25$~eV. (a) Spectrogram with initial state
polarization. (b) Spectrogram without initial state polarization.
(c) Corresponding centers of energies $\delta E_{COE}(\tau)$.
\label{fig:fig5} } \vspace{-6mm}
\end{center}
\end{figure}

\begin{figure}[t]
\begin{center}
\includegraphics[width=1.0\columnwidth,keepaspectratio=true,
draft=false]{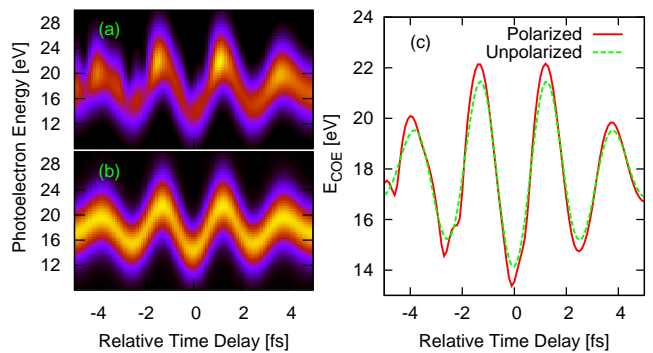}  \vspace{-6mm} \caption{(Color online) Same
as Fig.~\ref{fig:fig5}, but for the first excited state of the 1d
hydrogen atom. \label{fig:fig6} } \vspace{-6mm}
\end{center}
\end{figure}

We first consider the non-degenerate case. For our one-dimensional
hydrogen atom, all levels are non-degenerate. In our TDSE
calculation, the initial-state polarization by the ir pulse can be
included (excluded) by keeping (dropping) the term $xE_L(t)$ in
(\ref{eq:init}). In Figs.~\ref{fig:fig5} and \ref{fig:fig6} we
compare the polarized (a) and unpolarized (b) spectrograms and their
corresponding centers of energy (c) for the ground state level and
the first excited state. Due to its large separation in energy from
all excited states, the effect of the polarization of the ground
state by the laser pulse on the spectrum is small. It slightly
increases the oscillation amplitude but barely changes the temporal
shift $\delta\tau$. In contrast, the first excited state, whose
binding energy is 6.34~eV, can be easily polarized due to its
laser-induced coupling to the second excited level at 3.64~eV. As
can be seen in Fig.~\ref{fig:fig6}, the polarization distorts the
spectrogram for negative delays where the ir pulse precedes the xuv
pulse. This distortion does not uniformly shift the spectrograms.
Therefore, $\delta\tau$ cannot be uniquely defined as a
delay-independent temporal shift.

For the degenerate case, we consider a space spanned by the
stationary wavefunctions $\psi_{200}(\br)$ and $\psi_{210}(\br)$ of
the real (3-dimensional) hydrogen atom. Under the influence of the
laser pulse, the wavefunction is
\begin{align}
\psi(\br,t)=\left[a_{200}(t)\psi_{200}(\br)
+a_{210}(t)\psi_{210}(\br)\right].
\end{align}
By shifting the energy scale such that the binding energies of the
two degenerate stationary states are
$\varepsilon_{200}=\varepsilon_{210}=0$, and by substituting
$\psi(\br,t)$ into the TDSE
\begin{align}
i\frac{\partial}{\partial
t}\psi(\br,t)=\left[H_{at}+zE_{L}(t)\right]\psi(\br,t),
\end{align}
we obtain the equations of motion for the coefficients $a_{200}(t)$
and $a_{210}(t)$,
\begin{align}
i\frac{d}{dt}a_{200}(t)&=\mu E_L(t)a_{210}(t),\\
i\frac{d}{dt}a_{210}(t)&=\mu E_L(t)a_{200}(t),
\end{align}
where $H_{at}$ is the atomic Hamiltonian and
$\mu=\langle\psi_{200}(\br)|z|\psi_{210}(\br)\rangle$ the
dipole-coupling matrix element. The above equations can be solved
analytically~\cite{Grossmann08},
\begin{align}
a_{200}(t)&=a_{200}^0\cos\left[\mu
A_L(t)\right]+ia_{210}^0\sin\left[\mu A_L(t)\right],\\
a_{210}(t)&=a_{210}^0\cos\left[\mu
A_L(t)\right]+ia_{200}^0\sin\left[\mu A_L(t)\right],
\end{align}
where $a_{200}^0$ and $a_{210}^0$ are the initial amplitudes at
$t_0\rightarrow-\infty$.

For example, the initial values $a_{200}^0=\pm1/\sqrt{2}$ and
$a_{210}^0=1/\sqrt{2}$ give the wavefunctions
\begin{align}
\label{eq:pm} \psi_{\pm}(\br,t)=\psi_{\pm}(\br)e^{\pm i\mu A_L(t)},
\end{align}
which evolve from the Stark states
\begin{align}
\psi_{\pm}(\br)=\frac{1}{\sqrt{2}}\left[\psi_{210}(\br)\pm\psi_{200}(\br)\right].
\end{align}
Similarly, for $a_{200}^0=1$ and $a_{210}^0=0$, we obtain the
wavefunction
\begin{align}
\label{eq:2s}
\psi_{2s}(\br,t)=\frac{1}{\sqrt{2}}\left[\psi_+(\br)e^{i\mu
A_L(t)}-\psi_-(\br)e^{-i\mu A_L(t)}\right],
\end{align}
which evolves from an initial 2s state, while $a_{200}^0=0$ and
$a_{210}^0=1$ results in a wavefunction that evolves from an
stationary 2p state,
\begin{align}
\label{eq:2p}
\psi_{2p}(\br,t)=\frac{1}{\sqrt{2}}\left[\psi_+(\br)e^{i\mu
A_L(t)}+\psi_-(\br)e^{-i\mu A_L(t)}\right].
\end{align}

\begin{figure}[t]
\begin{center}
\includegraphics[width=1.0\columnwidth,keepaspectratio=true,
draft=false]{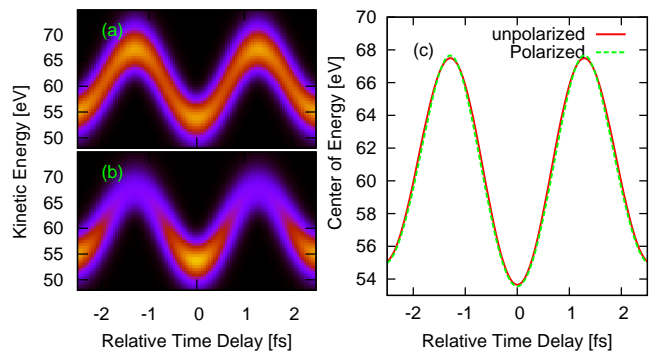}  \vspace{-6mm} \caption{(Color online)
Photoelectron spectrum for an initial 2s state of hydrogen:
(a) Neglecting ir-laser-induced initial-state polarization by
setting $\mu=0$ in (\ref{eq:2s}).
(b) Including laser-induced initial-state polarization, using
$\mu=3$~a.u. in (\ref{eq:2s}).
(c) Corresponding centers of energies $E_{COE}(\tau)$. No
polarization-induced temporal shift is observed between the results
for polarized and unpolarized initial states. Similar results (not
shown) were obtained for initial 2p states, using (\ref{eq:2p}).
\label{fig:fig7} } \vspace{-6mm}
\end{center}
\end{figure}

Next, we calculate the ir-streaked spectrum using any of the
wavefunctions (\ref{eq:pm})-(\ref{eq:2p}) as the initial state in
(\ref{eq:tmatrix}). In order to disentangle temporal shifts induced
by i) the initial-state polarization (relative to an unpolarized
target) and ii) the Coulomb potential acting on the final PE state
(relative to the SFA, see section III), we neglect the final-state
distortion by the Coulomb potential and study initial-state
polarization effects within the SFA. In this polarization-effect
study, we hence use the Volkov wavefunction $\psi_{\bk}(\br,t)$ as
an approximation to the final state.
If the initial state evolves from a stationary 2s or 2p state
according to (\ref{eq:2s}) or (\ref{eq:2p}), the polarization causes
delay-dependent interferences between the two Stark states
$\psi_{\pm}$. This interference significantly changes the
energy-differential PE yield in the streaking trace in
Fig.~\ref{fig:fig7} b relative to the trace for an unpolarized
initial state in Fig.~\ref{fig:fig7} a. However, the interference
does not induce a relative temporal shift of the polarized relative
to the unpolarized spectrum, which is best seen in the centers of
energy of the two spectra in Fig.~\ref{fig:fig7} c.
This lack of an interference-induced temporal shift is explained by
the fact that the dipole expectation values
$\langle\psi_{2s}(\br,t)|z|\psi_{2s}(\br,t)\rangle$ and
$\langle\psi_{2p}(\br,t)|z|\psi_{2p}(\br,t)\rangle$ are zero at all
times, even though the ir-laser pulse mixes the stationary 2s and 2p
states.

\begin{figure}[b]
\begin{center}
\includegraphics[width=1.0\columnwidth,keepaspectratio=true,
draft=false]{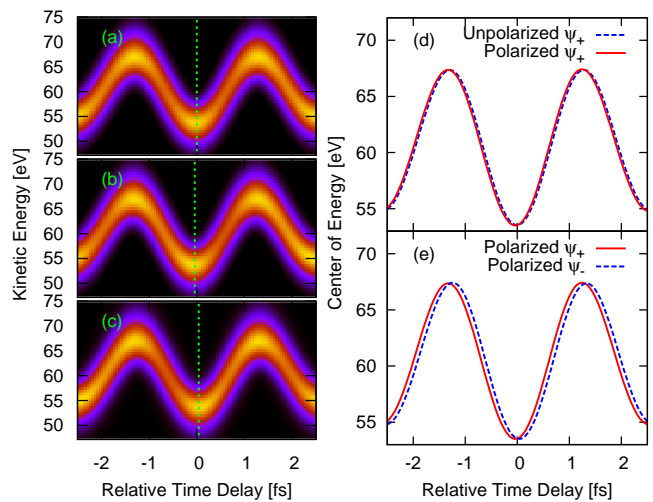}  \vspace{-6mm} \caption{(Color online)
Photoelectron spectra for the initial n=2 Stark states (\ref{eq:pm})
of hydrogen:
(a) For $\psi_+(\br,t)$, neglecting ir-laser-induced initial-state
polarization by setting $\mu=0$ in (\ref{eq:pm}).
(b) For $\psi_+(\br,t)$, including laser-induced initial-state
polarization, using $\mu=3$~a.u. in (\ref{eq:pm}).
(c) For $\psi_-(\br,t)$, including laser-induced initial-state
polarization, using $\mu=3$~a.u. in (\ref{eq:pm}).
(d,e) Corresponding centers of energies $ E_{COE}(\tau)$, showing a
relative temporal shift  between the streaking traces of (d)
polarized and unpolarized initial $\psi_+(\br,t)$ states and (e)
polarized initial $\psi_+(\br,t)$ and $\psi_-(\br,t)$ states.
 \label{fig:fig8} } \vspace{-6mm}
\end{center}
\end{figure}

The situation is different for the states (\ref{eq:pm}) that evolve
out of initial Stark states $\psi_{\pm}$. The comparison of the
energy-differential PE yields in the streaking trace for initial
states (\ref{eq:pm}) with and without including ir-laser-induced
initial-state polarization shows only very small, hardly noticeable,
differences (Fig.s~\ref{fig:fig8} (a-c)). However, temporal
shifts~\cite{Baggesen10} become noticeable in the corresponding
centers of energies. At a PE energy of 60~eV, they amount to 41~as
between an ir-laser polarized and unpolarized initial $\psi_{+}$
state (Fig.~\ref{fig:fig8} (d)) and to  82~as between polarized
initial $\psi_{+}$ and $\psi_{-}$ states (Fig.~\ref{fig:fig8} (e)).
These shifts originate in the permanent dipole moments of the Stark
states whose interaction with the ir-laser electric field shifts the
streaked spectra.

\section{Conclusions}
\label{sec:conclusion}

We have shown how the simultaneous interaction of an xuv PE with the
electric field of a streaking ir laser pulse and the Coulomb
potential of the residual ion induces a specific
Coulomb-Laser-coupling phase and leads to an attosecond temporal
shift and amplitude enhancement in the oscillation of the streaked
PE spectrum. This shift and amplitude enhancement become significant
and observable as the xuv photon energy approaches the ionization
threshold.  It can be explained semiclassically in terms of an added
Coulomb-phase factor in the PE wave function. This factor reveals
the origin of the observable temporal shift as a Coulomb-laser
coupling effect in the PE dynamics: the PE absorbs and releases ir
photons while moving subject to the ionic Coulomb force. The
analytical results obtained in  EA show that the CL coupling induces
a temporal shift relative to $A_L$, thus relative to the SFA result.
For the experimental observation of $\delta\tau$ and $K/k$ as a
function of the PE kinetic energy, we suggest using xuv pulses with
tunable xuv photon energy to photoemit electrons from two levels
with a large energy separation~\cite{Schultze10}.

We have also examined the effect of ir-laser-induced polarization of
the initial state on the ir-streaked xuv PE spectrum. If the initial
state is not degenerate and has a large energetic separation from
all other states, its very small polarization does not noticeably
affect the PE spectrum. On the other hand, if the initial state can
easily be coupled to other states by the ir-laser pulse, its
polarization is important and, interestingly, does not uniformly
shift the spectrum. If the initial state  has a permanent dipole
moment, such as the n=2 Stark states of hydrogen, there is a
relative temporal shift in the streaking traces i) for different
initial Stark states and ii) with and without inclusion of the
initial-state polarization.

\begin{acknowledgments}
We thank F. He for helpful discussions. This work was supported by
the NSF and the Division of Chemical Sciences, Office of Basic
Energy Sciences, Office of Energy Research, US~DOE. Some of the
numerical computations for this project were performed on the Beocat
cluster at Kansas State University.
\end{acknowledgments}

\begin{widetext}
\appendix

\section{Using Eikonal Wavefunction for RABITT}
\label{sec:appendix_A}

In this appendix, we show that atomic phase in RABITT and the
relative temporal shift induced by the Coulomb interaction are
identical within an eikonal approximation. We start from the ir
assisted single xuv-photon photoemission amplitude
\begin{align}
\label{eq:Tfi}
T_{fi}(\tau)=-i\int_{-\infty}^{+\infty}dt\langle\psi_{f}(t)|\br\cdot
E_{X}(t+\tau)|\psi_i(t)\rangle
\end{align}
where $\psi_{f}(t)$ and $\psi_{i}(t)$ are ir-dressed final and
initial states, respectively. Different from streaking, in RABITT
attosecond pulse trains (APT) synthesized from a number of odd
harmonics of the ir field
\begin{align} \label{eq:apt}
E_X(t)=E_{X,0}\sum_{n}e^{-i\omega_nt+i\varphi_n}
\end{align}
are used. Here $\omega_n=(2n+1)\omega_L$ and $\varphi_n$ are the
frequency and phase of the $(2n+1)$-th  harmonic, respectively, and
$\omega_L$ is the fundamental frequency of the ir field. For
simplicity, all harmonics are assumed to have the same strength
$E_{X,0}$. Using these APT, we obtain a series of peaks in the PE
spectrum that are separated by twice the ir photon energy. In the
presence of a weak ir field, sidebands will form between the main
peaks due to the emission or absorption of ir photons. In RABITT
periodic intensity variations are observed in the first sideband due
to the inteference of two distinct two-photon transition routes: 1)
absorption of one harmonic photon with frequency
$\omega_{n}=(2n+1)\omega_L$ and emission of an ir photon with
frequency $\omega_L$, 2) absorption of an adjacent lower harmonic
photon with frequency $\omega_{n-1}=(2n-1)\omega_L$ and an ir photon
with frequency $\omega_L$. The intensity of the sideband is
controlled by the delay $\tau$ between the APT and the ir field,
\begin{align}
P_{sb}(\tau)\sim\left[1-\cos\left(2\omega_L\tau+\Delta\varphi_{n}-\Delta\phi^{at}\right)
\right].
\end{align}
It is shifted by the harmonic phases
$\Delta\varphi_{n}=\varphi_{n-1}-\varphi_{n}$ and the {\em atomic
phase} $\Delta\phi^{at}$~\cite{Veniard96,Paul01,Toma02}.
$\Delta\phi^{at}$ is a function of the PE energy.

Using the eikonal approximated wavefunction, we can now show that
the atomic phase $\Delta\phi^{at}/2\omega_L$ is equal to the
temporal shift $\delta\tau$ in streaked spectra. Using Eq.~
(\ref{eq:Gk2}) in the main text for the EA phase $S^{EA}_k$,
expanding $\psi^{EA}_{k}(x,t)$ up to first order in $E_L$,
\begin{align}
\psi^{EA}_{k}(x,t)=\frac{1}{(2\pi)^{1/2}}e^{i(k+A_L(t))x}e^{-iE_kt+iS_C(x)}
\left[1-\frac{iKE_L(t+\delta\tau)}{\omega_L^2}\right],
\end{align}
substituting it into (\ref{eq:Tfi}) for $\psi_{f}$, and carrying out
the time integration, we obtain the transition amplitude up to
two-photon process
\begin{align}
T(\tau) =&-id_kE_{X,0}\sum_n\delta\left[E+I_p-(2n+1)\omega_L\right]
\nonumber\\
&-i\frac{Kd_k}{\omega_L^2}\frac{E_{L,0}E_{X,0}}{2}\sum_n
\left\{\delta(E_k+I_p-2n\omega_L)e^{-i\omega_L(\tau-\delta\tau)}-
\delta\left[E_k+I_p-(2n+2)\omega_L\right]e^{+i\omega_L(\tau-\delta\tau)}\right\},
\end{align}
where $E_k=k^2/2$ is the PE kinetic energy, $e^{-i(E_k+I_p)\tau}$
has been dropped,
\begin{align}
d_k&=\frac{1}{(2\pi)^{1/2}}\int
dxe^{-i(k+A_L(t))x}xe^{-iS_C(x)}\psi_i(x),
\end{align}
and
\begin{align}
\tilde{E}_L(\omega)&=\frac{E_{L,0}}{2i}
\left[\delta(\omega+\omega_L)-\delta(\omega-\omega_L)\right],\\
\tilde{E}_X(\omega)&=\sum_{n}E_{X,0}e^{+i\varphi_n}\delta[\omega-(2n+1)\omega_L]
\end{align}
are the Fourier transformations of $E_L(t)$ and $E_X(t)$. The
transition amplitude for sidebands at PE energies
$E_k=2n\omega_L-I_p$ follows as
\begin{align}
T_{sb}(E_f,\tau)
=&-i\frac{Kd_kE_{L,0}}{2\omega_L^2}E_{X,0}e^{-i\varphi_n-i\omega_L(\tau-\delta\tau)}
\left[1-e^{i\Delta\varphi_n+2i\omega_L(\tau-\delta\tau)}\right].
\end{align}
Accordingly, the sideband intensity as a function of delay $\tau$ is
\begin{align}
P_{sb}(E_k,\tau)=\left|T_{sb}(\tau)\right|^2=\frac{K^2d^2_kE_{L,0}^2E_{X,0}^2}{2\omega_L^4}
\left\{1-\cos\left[2\omega_L(\tau-\delta\tau)+\Delta\varphi_n\right]\right\}
\end{align}
from which the atomic phase $\Delta\phi^{at}/2\omega_L=\delta\tau$
is identified.

\end{widetext}

\bibliographystyle{apsrev}
\bibliography{attosecond_con}

\begin{thebibliography}{24}
\expandafter\ifx\csname natexlab\endcsname\relax\def\natexlab#1{#1}\fi
\expandafter\ifx\csname bibnamefont\endcsname\relax
  \def\bibnamefont#1{#1}\fi
\expandafter\ifx\csname bibfnamefont\endcsname\relax
  \def\bibfnamefont#1{#1}\fi
\expandafter\ifx\csname citenamefont\endcsname\relax
  \def\citenamefont#1{#1}\fi
\expandafter\ifx\csname url\endcsname\relax
  \def\url#1{\texttt{#1}}\fi
\expandafter\ifx\csname urlprefix\endcsname\relax\def\urlprefix{URL }\fi
\providecommand{\bibinfo}[2]{#2}
\providecommand{\eprint}[2][]{\url{#2}}

\bibitem[{Kra()}]{Krausz09}
\bibinfo{note}{F. Krausz and M. Ivanov, Rev. Mod. Phys. {\bf 81}, 163 (2009),
  and ref.s therein}.

\bibitem[{Lew()}]{Lewenstein94}
\bibinfo{note}{M. Lewenstein, Ph. Balcou, M. Yu. Ivanov, Anne L'Huillier, and
  P. B. Corkum, Phys. Rev. A {\bf 49}, 2117 (1994)}.

\bibitem[{Zha()}]{Zhang09}
\bibinfo{note}{C.-H. Zhang and U. Thumm, Phys. Rev. Lett. {\bf 102}, 123601
  (2009)}.

\bibitem[{Cav()}]{Cavalieri07}
\bibinfo{note}{A. L. Cavalieri, N. M\"{u}ller, Th. Uphues, V. S. Yakovlev, A.
  Baltus caronka, B. Horvath, B. Schmidt, L. Bl\"{u}mel, R. Holzwarth, S.
  Hendel, M. Drescher, U. Kleineberg, P. M. Echenique, R. Kienberger, F.
  Krausz, and U. Heinzmann, Nature {\bf 449}, 1029 (2007)}.

\bibitem[{Kaz({\natexlab{a}})}]{Kazansky09}
\bibinfo{note}{A. K. Kazansky and P. M. Echenique, Phys. Rev. Lett. {\bf 102}
  177401 (2009)}.

\bibitem[{pri()}]{private}
\bibinfo{note}{We thank A. L. Cavalieri, and N. Karpowicz for sharing their
  unpublished experimental data.}

\bibitem[{Var({\natexlab{a}})}]{Varro98}
\bibinfo{note}{S. Varr\'{o} and F. Ehlotzky, J. Phys. B {\bf 31}, 2145 (1998)}.

\bibitem[{Lem()}]{Lemell09}
\bibinfo{note}{C. Lemell, B. Solleder, K. T\"{o}k\'{e}si, and J.
  Burgd\"{o}rfer, Phys. Rev. A {\bf 79}, 062901 (2009)}.

\bibitem[{Bag({\natexlab{a}})}]{Baggesen09}
\bibinfo{note}{J. C. Baggesen and L. B. Madsen, Phys. Rev. A {\bf 78} 032903
  (2008); ibid. {\bf 80} 030901(R) (2009)}.

\bibitem[{Kro()}]{Kroll73}
\bibinfo{note}{N. M. Kroll and K. M. Watson, Phys. Rev. A {\bf 8}, 804 (1973)}.

\bibitem[{Smi({\natexlab{a}})}]{Smirnova08}
\bibinfo{note}{O. Smirnova, M. Spanner, and M. Ivanov, Phys. Rev. A {\bf 77}
  033407 (2008); J. Phys. B {\bf 40}, F197 (2007)}.

\bibitem[{Ven()}]{Veniard96}
\bibinfo{note}{V. Veniard, R. Ta\"{\i}eb, and A. Maquet, Phys. Rev. A {\bf 54},
  721 (1996))}.

\bibitem[{Tom()}]{Toma02}
\bibinfo{note}{E. S. Toma and H. G. Muller, J. Phys. B {\bf 35}, 3435 (2002)}.

\bibitem[{Mau()}]{Mauritsson05}
\bibinfo{note}{J. Mauritsson, M. B. Gaarde, and K. J. Schafer, Phys. Rev. A
  {\bf 72}, 013401 (2005)}.

\bibitem[{Var({\natexlab{b}})}]{Varju05}
\bibinfo{note}{K. Varj\'{u}, P. Johnsson, R. L\'{o}pez-Martens, T. Remetter, E.
  Gustafsson, J. Mauritsson, M. B. Gaarde, K. J. Schafer, Ch. Erny, I. Sola, A.
  Za\"{\i}r, E. Constant, E. Cormier, E. M\'{e}vel, and A. L'Huillier, Laser
  Physics {\bf 15}, 888 (2005)}.

\bibitem[{\citenamefont{Joachain}(1983)}]{Joachain83}
\bibinfo{author}{\bibfnamefont{C.}~\bibnamefont{Joachain}},
  \emph{\bibinfo{title}{Quantum Collision Theory}} (\bibinfo{address}{New
  York}, \bibinfo{year}{1983}).

\bibitem[{\citenamefont{Gersten and Mittleman}(1975)}]{Gersten75}
\bibinfo{author}{\bibfnamefont{J.~I.} \bibnamefont{Gersten}} \bibnamefont{and}
  \bibinfo{author}{\bibfnamefont{M.~H.} \bibnamefont{Mittleman}},
  \bibinfo{journal}{Phys. Rev. A} \textbf{\bibinfo{volume}{12}},
  \bibinfo{pages}{1840} (\bibinfo{year}{1975}).

\bibitem[{Thu()}]{Thumm92}
\bibinfo{note}{U. Thumm, J. Phys. B {\bf 25}, 421 (1992), and ref.s therein}.

\bibitem[{Kaz({\natexlab{b}})}]{Kazansky07}
\bibinfo{note}{A. K. Kazansky and N. M. Kabachnik, J. Phys. B {\bf 40} 2163
  (2007)}.

\bibitem[{Smi({\natexlab{b}})}]{Smirnova06}
\bibinfo{note}{O. Smirnova, A S Mouritzen, S Patchkovskii, and M. Ivanov, J.
  Phys. B {\bf 39}, S323 (2006)}.

\bibitem[{Bag({\natexlab{b}})}]{Baggesen10}
\bibinfo{note}{J. C. Baggesen and L. B. Madsen, Phys. Rev. Lett. {\bf 104}
  043602 (2010)}.

\bibitem[{\citenamefont{Grossmann}(2008)}]{Grossmann08}
\bibinfo{author}{\bibfnamefont{F.}~\bibnamefont{Grossmann}},
  \emph{\bibinfo{title}{Theoretical Femtosecond Physics}}
  (\bibinfo{publisher}{Springer}, \bibinfo{address}{Berlin},
  \bibinfo{year}{2008}).

\bibitem[{Sch()}]{Schultze10}
\bibinfo{note}{M. Schultze, M. Fie{\ss}, N. Karpowicz, J. Gagnon, M. Korbman,
  M. Hofstetter, S. Neppl, A. L. Cavalieri, Y. Komninos, Th. Mercouris, C. A.
  Nicolaides, R. Pazourek, S. Nagele, J. Feist, J. Burgd\"{o}rfer, A. M.
  Azzeer, R. Ernstorfer, R. Kienberger, U. Kleineberg, E. Goulielmakis, F.
  Krausz, and V. S. Yakovlev, Science {\bf 328}, 1658 (2010)}.

\bibitem[{Pau()}]{Paul01}
\bibinfo{note}{P. M. Paul E. S. Toma, P. Breger, G. Mullot, F. Aug\'{e}, Ph.
  Balcou, H. G. Muller, and P. Agostini, Science {\bf 292}, 1689 (2001)}.

\end{thebibliography}

\end{document}